# Computation of time-optimal control problem with variation evolution principle

Sheng ZHANG and Wei-Qi QIAN

(2017.03)

*Abstract:* **An effective form of the Variation Evolving Method (VEM), which originates from the continuous-time dynamics stability theory, is developed for the classic time-optimal control problem with control constraint. Within the mathematic derivation, the Pontryagin's Minimum Principle (PMP) optimality conditions are used. Techniques including limited integrator and corner points are introduced to capture the right solution. The variation dynamic evolving equation may be reformulated as the Partial Differential Equation (PDE), and then discretized as finite-dimensional Initial-value Problem (IVP) to be solved with common Ordinary Differential Equation (ODE) integration methods. An illustrative example is solved to show the effectiveness of the method. In particular, the VEM is further developed to be more flexible in treating the boundary conditions of the Optimal Control Problem (OCP), by initializing the transformed IVP with arbitrary initial values of variables.**

## I. INTRODUCTION

Optimal control theory aims to determine the inputs to a dynamic system that optimize a specified performance index while satisfying constraints on the motion of the system. It is closely related to engineering and has been widely studied [1]. Because of the complexity, usually Optimal Control Problems (OCPs) are solved with numerical methods. Various numerical methods are developed and generally they are divided into two classes, namely, the direct methods and the indirect methods [2]. The direct methods discretize the control or/and state variables to obtain the Nonlinear Programming (NLP) problem, for example, the widely-used direct shooting method [3] and the classic collocation method [4]. These methods are easy to apply, whereas the results obtained are usually suboptimal [5], and the optimal may be infinitely approached. The indirect methods transform the OCP to a Boundary-value Problem (BVP) through the optimality conditions. Typical methods of this type include the well-known indirect shooting method [2] and the novel symplectic method [6]. Although be more precise, the indirect methods often suffer from the significant numerical difficulty due to ill-conditioning of the Hamiltonian dynamics, that is, the stability of costates dynamics is adverse to that of the state dynamics [7]. The recent development, representatively the Pseudo-spectral (PS) method [8], blends the two types of methods, as it unifies the NLP and the BVP in a dualization view [9]. Such methods inherit the advantages of both types and blur their difference.

Theories in the control field often enlighten strategies for the optimal control computation, for example, the non-linear variable transformation to reduce the variables [10]. Recently, a new Variation Evolving Method (VEM), which is enlightened by the states evolving within the stable continuous-time dynamic system, is proposed for the optimal control computation [11]. It also

The authors are with the Computational Aerodynamics Institution, China Aerodynamics Research and Development Center, Mianyang, 621000, China. (e-mail: zszhangshengzs@hotmail.com).

synthesizes the direct and indirect methods, but from a new standpoint. In the VEM, the Partial Differential Equation (PDE), which describes the evolution of variables towards the extremal solution, is derived through the variation motion in typical OCPs. Using the well-known semi-discrete method in the field of PDE numerical calculation [12], the PDEs are transformed to the finite-dimensional Initial-value Problems (IVPs) to be solved, with the mature Ordinary Differential Equation (ODE) integration methods. Because the extremums are guaranteed be the equilibrium point of the deduced dynamic system, the optimal solution will be gradually approached. However, the strategy developed in Ref. [11] is not generally applicable to the state- and control-constrained OCPs, because the idea of constructing analogous equivalent functional is not available when complex path constraints are involved. Further studies along that thread may require employment of techniques such as the Karush–Kuhn–Tucker (KKT) variables or the slack variables [13].

In this paper, we restrict our scope to the classic time-optimal control problems with control constraint. Some explorative work is carried out, and an effective alternative, which also uses the variation evolution, is developed according to the Pontryagin's Minimum Principle (PMP) [14]. In our work, it is assumed that the solution for the optimization problem exists. We do not describe the existing conditions for the purpose of brevity. Relevant researches such as the Filippov-Cesari theorem are documented in [15]. In the following, first preliminaries that state the inspiration of the VEM are presented. Then the foundational VEM bred under this idea is recalled for the unconstrained calculus-of-variations problem. Next the computation of the time-optimal control problem with control constraint, from the variation evolution way, is investigated. Later an illustrative example is solved to verify the effectiveness of the proposed method.

## II. Preliminaries

The VEM is a newly developed method for the optimal solutions. To help understanding, its motivations are reviewed. For a continuous-time autonomous dynamic system like

$$\dot{\boldsymbol{x}} = \boldsymbol{f}(\boldsymbol{x}) \tag{1}$$

where $\boldsymbol{x} \in \mathbb{R}^n$ is the state, $\dot{\boldsymbol{x}} = \frac{\mathrm{d}\boldsymbol{x}}{\mathrm{d}t}$ is its time derivative, and $\boldsymbol{f} : \mathbb{R}^n \to \mathbb{R}^n$ is a vector function. Suppose that $\hat{\boldsymbol{x}}$ is a asymptotically stable equilibrium point of system (1) that satisfies $\boldsymbol{f}(\hat{\boldsymbol{x}}) = \boldsymbol{0}$, then from any initial condition $\boldsymbol{x}(t)\big|_{t=0} = \boldsymbol{x}_0$ within the stability domain $\mathbb{D}$ that contains $\hat{\boldsymbol{x}}$, the state $\boldsymbol{x}$ will tend to $\hat{\boldsymbol{x}}$ over time $t$ [16]. According to the Lyapunov theory, there is a continuously differentiable function $V : \mathbb{D} \to \mathbb{R}$ such that

i) $V(\hat{\boldsymbol{x}}) = 0$ and $V(\boldsymbol{x}) > 0$ in $\mathbb{D} / \{\hat{\boldsymbol{x}}\}$.

ii) $\dot{V}(\boldsymbol{x}) \le 0$

For example, maybe $\boldsymbol{f}(\boldsymbol{x})$ satisfies $(\boldsymbol{x} - \hat{\boldsymbol{x}})^{\mathrm{T}} \boldsymbol{f}(\boldsymbol{x}) \le 0$, and then a feasible Lyapunov function can be constructed as

$$V = \frac{1}{2}(\boldsymbol{x} - \hat{\boldsymbol{x}})^{\mathrm{T}}(\boldsymbol{x} - \hat{\boldsymbol{x}}) \tag{2}$$

The dynamics governed by $\boldsymbol{f}(\boldsymbol{x})$ determines that $\dot{V} \le 0$ and $\boldsymbol{x}$ will converge to the equilibrium $\hat{\boldsymbol{x}}$. Fig. 1 sketches the trajectory of some state in the stable dynamic system and the corresponding Lyapunov function value. No matter what the initial condition $\boldsymbol{x}_0$ is, as long as it falls into the stability domain $\mathbb{D}$, the state $\boldsymbol{x}$ will approaches the equilibrium $\hat{\boldsymbol{x}}$ gradually.

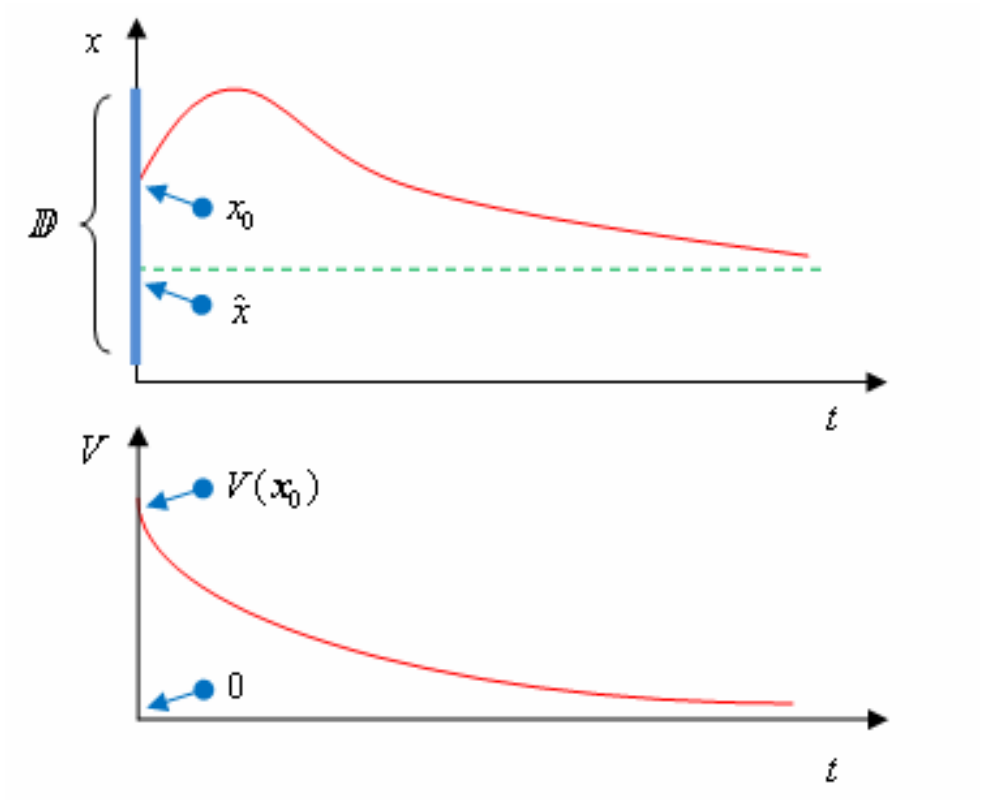

Fig. 1. The sketch for the state trajectory and the Lyapunov function value profile.

In the system dynamics theory, from the stable dynamics of state $\boldsymbol{x}$, we may construct a monotonously decreasing function $V(\boldsymbol{x})$, which will achieve its minimum when $\boldsymbol{x}$ reaches $\hat{\boldsymbol{x}}$. Inspired by it, now we consider its inverse problem, that is, from a performance index function to derive the dynamics that minimize this performance index. Consider the parameter optimization problem with performance index

$$J = h(\boldsymbol{\theta}) \tag{3}$$

where $\boldsymbol{\theta}$ is the optimization parameter vector and $h:\mathbb{R}^n \to \mathbb{R}$ is a scalar function. To find the optimal value $\hat{\boldsymbol{\theta}}$ that minimizes $J$, we make the analogy to the Lyapunov function and differentiate $J$, i.e., function $h$ here, with respect to a virtual time $\tau$, which is used to describe the derived dynamics.

$$\frac{\mathrm{d}J}{\mathrm{d}\tau} = \frac{\mathrm{d}h}{\mathrm{d}\tau} = {h_{\boldsymbol{\theta}}}^{\mathrm{T}} \frac{\mathrm{d}\boldsymbol{\theta}}{\mathrm{d}\tau} \tag{4}$$

where the column vector $h_{\boldsymbol{\theta}} = \dfrac{\partial h}{\partial \boldsymbol{\theta}}$ is the shorthand notation of the partial derivative, and the superscript " $\mathrm{T}$ " denotes the transpose operator. To guarantee that $J$ decreases with respect to $\tau$, i.e., $\dfrac{\delta J}{\delta \tau} \leq 0$, we may set

$$\frac{\mathrm{d}\boldsymbol{\theta}}{\mathrm{d}\tau} = -\boldsymbol{K} h_{\boldsymbol{\theta}} \tag{5}$$

where $\boldsymbol{K}$ is a positive definite matrix. Under this dynamics, $h$ will decrease until it reaches a extremum, and $\boldsymbol{\theta}$ will approaches $\hat{\boldsymbol{\theta}}$, the equilibrium point of system (5), which satisfies $h_{\boldsymbol{\theta}}\big|_{\boldsymbol{\theta}=\hat{\boldsymbol{\theta}}} = \boldsymbol{0}$. This equilibrium condition is exactly the first-order optimality condition for the optimization problem (3).

A bolder idea further arises hereafter. If the optimization parameter $\boldsymbol{\theta}$ can approach its optimal under the dynamics given by Eq. (5), we can imagine that a variable $\boldsymbol{x}(t)$ might also evolve to the optimal solution to minimize some performance index within certain dynamics. Fig. 2 illustrates the idea of the VEM in solving the OCP, and we formally introduce the virtual variation time $\tau$, a new dimension orthogonal to the normal time $t$, to describe the variation evolution process. Through the variation motion, the initial guess of variable will evolve to the optimal solution. In Ref. [11], the VEM bred under this idea is demonstrated for the unconstrained calculus-of-variations problems and the OCPs with dynamic constraint. We will consider the implementation for the time-optimal control problem with control constraint below.

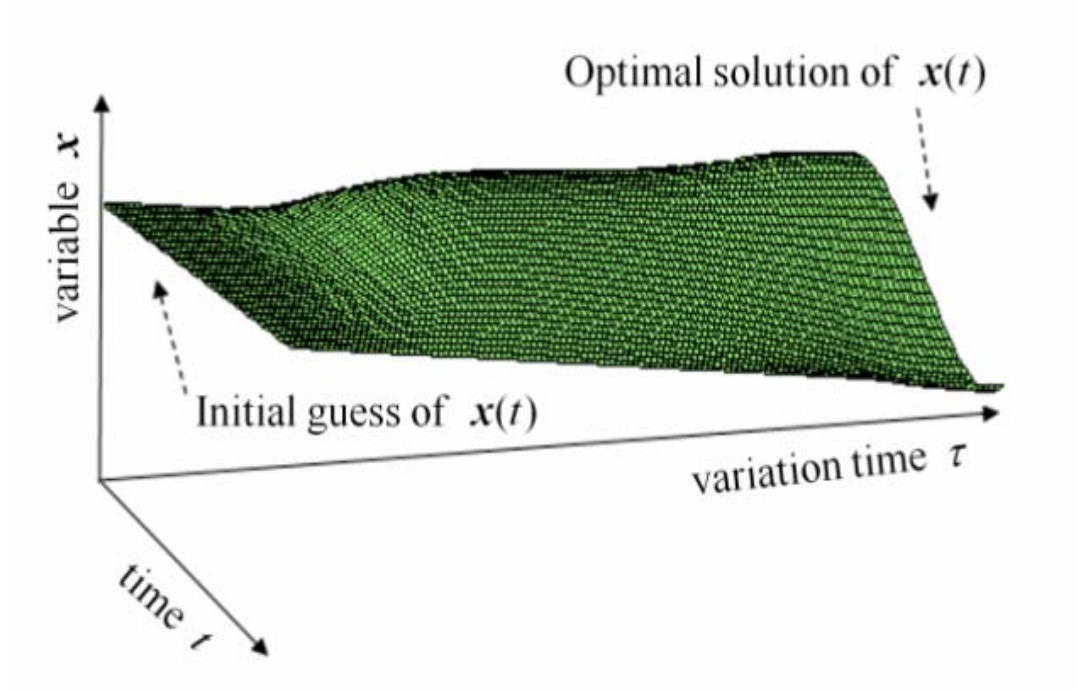

Fig. 2. The illustration of the variable evolving along the variation time $\tau$ in the VEM.

## III. The Foundational Variation Evolving Method

Before discussing the time-optimal control problem, the foundational VEM, which was first demonstrated in Ref. [11], is again presented for the unconstrained calculus-of-variations problem defined as

**Problem 1**: For the following functional depending on variable vector $\boldsymbol{y}(t) \in \mathbb{R}^n$

$$J = \int_{t_0}^{t_f} F\left(\boldsymbol{y}(t), \dot{\boldsymbol{y}}(t), t\right) \mathrm{d}t \tag{6}$$

where $t \in \mathbb{R}$ is the time. The elements of $\boldsymbol{y}$ belong to $C^2[t_0, t_f]$, which denotes the set of variables with continuous second-order derivatives (indicated by the superscript). The function $F : \mathbb{R}^n \times \mathbb{R}^n \times \mathbb{R} \to \mathbb{R}$ and its first-order and second-order partial derivatives are continuous with respect to $\boldsymbol{y}$, its time derivative $\dot{\boldsymbol{y}} = \frac{\mathrm{d}\boldsymbol{y}}{\mathrm{d}t}$ and $t$. $t_0$ and $t_f$ are the fixed initial and terminal time. Find the extremal solution $\hat{\boldsymbol{y}}$ that minimizes $J$, i.e.

$$\hat{\boldsymbol{y}} = \arg\min(J) \tag{7}$$

Follow the idea of dynamics evolution to reduce some performance index. We anticipate that any initial guess of $\boldsymbol{y}(t)$, whose elements belong to $C^2[t_0, t_f]$, will evolve to the minimum along the variation dimension. Like the decrease of a Lyapunov function, if $J$ in Eq. (6) decreases with respect to the variation time $\tau$, i.e., $\frac{\delta J}{\delta \tau} \le 0$, we may finally obtain the optimal solution. Differentiating (6) with respect to $\tau$ (even $\tau$ does not explicitly exist) produces

$$\begin{aligned} \frac{\delta J}{\delta \tau} &= \int_{t_0}^{t_f} \left( F_{\boldsymbol{y}}{}^{\mathrm{T}} \frac{\delta \boldsymbol{y}}{\delta \tau} + F_{\dot{\boldsymbol{y}}}{}^{\mathrm{T}} \frac{\delta \dot{\boldsymbol{y}}}{\delta \tau} \right) \mathrm{d}t \\ &= F_{\dot{\boldsymbol{y}}}{}^{\mathrm{T}} \left.\frac{\delta \boldsymbol{y}}{\delta \tau}\right|_{t_f} - F_{\dot{\boldsymbol{y}}}{}^{\mathrm{T}} \left.\frac{\delta \boldsymbol{y}}{\delta \tau}\right|_{t_0} + \int_{t_0}^{t_f} \left( \left[ F_{\boldsymbol{y}} - \frac{\mathrm{d}}{\mathrm{d}t}(F_{\dot{\boldsymbol{y}}}) \right]^{\mathrm{T}} \frac{\delta \boldsymbol{y}}{\delta \tau} \right) \mathrm{d}t \end{aligned} \tag{8}$$

From an arbitrary initial guess $\boldsymbol{y}(t)\big|_{\tau=0} = \tilde{\boldsymbol{y}}(t)$, then by enforcing $\frac{\delta J}{\delta \tau} \le 0$, we may set that

$$\frac{\delta \boldsymbol{y}}{\delta \tau} = -\boldsymbol{K}\left( F_{\boldsymbol{y}} - \frac{\mathrm{d}}{\mathrm{d}t}(F_{\dot{\boldsymbol{y}}}) \right), \quad t \in (t_0, t_f) \tag{9}$$

$$\frac{\delta \boldsymbol{y}(t_0)}{\delta \tau} = \boldsymbol{K} F_{\dot{\boldsymbol{y}}}(t_0) \tag{10}$$

$$\frac{\delta \boldsymbol{y}(t_f)}{\delta \tau} = -\boldsymbol{K} F_{\dot{\boldsymbol{y}}}(t_f) \tag{11}$$

where $\boldsymbol{K}$ is a $n \times n$ dimensional positive definite matrix. The variation dynamic evolving equations (9)-(11) describes the variation motion of $\boldsymbol{y}(t)$ starting from $\tilde{\boldsymbol{y}}(t)$, and it is proved that the motion is directed to the extremum [11]. It drives the performance index $J$ to decrease until $\frac{\delta J}{\delta \tau} = 0$, and when $\frac{\delta J}{\delta \tau}$ = 0, this determines the optimal conditions, namely, the Euler-Lagrange equation [17][18]

$$F_{\boldsymbol{y}} - \frac{\mathrm{d}}{\mathrm{d}t}(F_{\dot{\boldsymbol{y}}}) = \boldsymbol{0} \tag{12}$$

The variation dynamic evolving equation (9) may be considered from the view of PDE formulation, by replacing the variation operation " $\delta$ " and the differential operator " d " with the partial differential operator " $\partial$ " as

$$\frac{\partial \boldsymbol{y}}{\partial \tau} = -\boldsymbol{K}\left( F_{\boldsymbol{y}} - \frac{\partial}{\partial t}(F_{\dot{\boldsymbol{y}}}) \right) \tag{13}$$

For this PDE, its right function only depends on the time $t$. Thus it is suitable to be solved with the semi-discrete method in the field of PDE numerical calculation. With the discretization along the normal time dimension, Eq. (13) is transformed to be IVPs with finite states. Note that the resulting IVP is defined with respect to the variation time $\tau$, not the normal time $t$. In the previous work [11], a demonstrative example is solved to verify the result.

## IV. Time-optimal control problems with control constraint

Consider the classic time-optimal control problem with control constraint.

**Problem 2**: Consider the time performance index

$$J_1 = t_f \tag{14}$$

subject to the autonomous dynamic equation

$$\dot{\boldsymbol{x}} = \boldsymbol{f}(\boldsymbol{x}, \boldsymbol{u}) \tag{15}$$

and control constraint

$$u_{\min} \le u_i \le u_{\max},\ i = 1, 2, ..., m \tag{16}$$

where $\boldsymbol{x} \in \mathbb{R}^n$ are the states and each element is piecewise second-order differentiable. $\boldsymbol{u} \in \mathbb{R}^m$ are the control inputs and each element is piecewise first-order differentiable. The vector function $\boldsymbol{f}: \mathbb{R}^n \times \mathbb{R}^m \to \mathbb{R}^n$ and its first-order and second-order partial derivatives are continuous and Lipschitz in $\boldsymbol{x}$ and $\boldsymbol{u}$. $u_{\min}$ and $u_{\max}$ are the lower and upper control limit, respectively. The initial boundary conditions at the fixed initial time $t_0$ and the terminal boundary conditions are

$$\boldsymbol{x}(t_0) = \boldsymbol{x}_0 \tag{17}$$

$$\boldsymbol{x}(t_f) = \boldsymbol{x}_f \tag{18}$$

Find the extremal solution $(\hat{\boldsymbol{x}}, \hat{\boldsymbol{u}})$ that minimizes $J_1$, i.e.

$$(\hat{\boldsymbol{x}}, \hat{\boldsymbol{u}}) = \arg\min(J) \tag{19}$$

From the PMP, the optimality conditions of Problem 2 are the state-costate differential equations

$$\dot{\boldsymbol{x}} - H_{\boldsymbol{\lambda}} = \boldsymbol{0} \tag{20}$$

$$\dot{\boldsymbol{\lambda}} + H_{\boldsymbol{x}} = \boldsymbol{0} \tag{21}$$

and the control algebraic equation

$$\boldsymbol{u} = \arg \min_{u_{\min} \le u_i \le u_{\max}} (H) \tag{22}$$

The transversality condition regarding the terminal time $t_f$ is

$$H(t_f) + 1 = 0 \tag{23}$$

where $\lambda$ is the costate vector and $H = \boldsymbol{\lambda}^{\mathrm{T}} \boldsymbol{f}$ is the Hamiltonian. Because the dynamic system is autonomous, we may obtain

$$\frac{\mathrm{d}H}{\mathrm{d}t} = 0 \tag{24}$$

which means

$$H(t) = -1 \tag{25}$$

In Ref. [11], equivalent unconstrained functional problem that has the same extremum as the OCP with dynamic constraint is constructed, with the first-order optimality conditions. Here for the time-optimal control problem defined in Problem 2, we will formulate a functional optimization problem, whose minimum satisfies the aforementioned optimality conditions (20), (21) and (25). Especially, we modified the unconstrained functional to effectively accommodate the boundary conditions. Thus the prescribed initial condition for variables at $(t = t_0, \tau = 0)$ or $(t = t_f, \tau = 0)$, as in Ref. [11], may be avoided.

**Problem 3**: Consider the following unconstrained functional

$$\begin{aligned} J_2 = & \left(\boldsymbol{x}(t_0) - \boldsymbol{x}_0\right)^{\mathrm{T}} \boldsymbol{W}_{\boldsymbol{x}_0} \left(\boldsymbol{x}(t_0) - \boldsymbol{x}_0\right) + \left(\boldsymbol{x}(t_f) - \boldsymbol{x}_f\right)^{\mathrm{T}} \boldsymbol{W}_{\boldsymbol{x}_f} \left(\boldsymbol{x}(t_f) - \boldsymbol{x}_f\right) \\ & + \int_{t_0}^{t_f} \left\{ \left(\dot{\boldsymbol{x}} - H_{\boldsymbol{\lambda}}\right)^{\mathrm{T}} \left(\dot{\boldsymbol{x}} - H_{\boldsymbol{\lambda}}\right) + \left(\dot{\boldsymbol{\lambda}} + H_{\boldsymbol{x}}\right)^{\mathrm{T}} \left(\dot{\boldsymbol{\lambda}} + H_{\boldsymbol{x}}\right) + (H+1)^2 \right\} \mathrm{d}t \end{aligned} \tag{26}$$

where $\boldsymbol{x}, \boldsymbol{\lambda} \in \mathbb{R}^n$ and their elements are piecewise second-order differentiable, $\boldsymbol{u} \in \mathbb{R}^m$ is piecewise first-order differentiable. The Hamiltonian $H$ and its first-order and second-order partial derivatives, with respect to $\boldsymbol{x}$, $\boldsymbol{\lambda}$, and $\boldsymbol{u}$, are piecewise continuous. The initial time $t_0$ is fixed and the terminal time $t_f$ is free. $\boldsymbol{W}_{\boldsymbol{x}_0}$ and $\boldsymbol{W}_{\boldsymbol{x}_f}$ are weighted positive definite matrixes. Find the extremal solution $(\hat{\boldsymbol{x}}, \hat{\boldsymbol{\lambda}}, \hat{\boldsymbol{u}})$ that minimizes $J_2$, i.e.

$$(\hat{\boldsymbol{x}}, \hat{\boldsymbol{\lambda}}, \hat{\boldsymbol{u}}) = \arg\min(J_2) \tag{27}$$

Replacing the function and variables in (6) respectively with

$$F = \left(\dot{\boldsymbol{x}} - H_{\boldsymbol{\lambda}}\right)^{\mathrm{T}} \left(\dot{\boldsymbol{x}} - H_{\boldsymbol{\lambda}}\right) + \left(\dot{\boldsymbol{\lambda}} + H_{\boldsymbol{x}}\right)^{\mathrm{T}} \left(\dot{\boldsymbol{\lambda}} + H_{\boldsymbol{x}}\right) + (H+1)^2 \tag{28}$$

$$\boldsymbol{y} = \begin{bmatrix} \boldsymbol{x} \\ \lambda \\ \boldsymbol{u} \end{bmatrix} \tag{29}$$

From the principle of the VEM and with extra consideration on the free terminal time, we may deduce the variation dynamic evolving equations as

$$\frac{\delta \boldsymbol{y}}{\delta \tau} = -2\boldsymbol{K}\boldsymbol{p}, \quad t \in (t_0, t_f) \tag{30}$$

and the boundary value variation dynamic equations as

$$\frac{\delta \boldsymbol{y}(t_0)}{\delta \tau} = 2\boldsymbol{K}\left.\begin{bmatrix} \boldsymbol{W}_{\boldsymbol{x}_0}\left(\boldsymbol{x}_0 - \boldsymbol{x}(t_0)\right) + \left(\dot{\boldsymbol{x}} - H_{\boldsymbol{\lambda}}\right) \\ \left(\dot{\boldsymbol{\lambda}} + H_{\boldsymbol{x}}\right) \\ \boldsymbol{p}_{\boldsymbol{u}} \end{bmatrix}\right|_{t_0} \tag{31}$$

$$\frac{\delta \boldsymbol{y}_f}{\delta \tau} = -2\boldsymbol{K}\left.\begin{bmatrix} \boldsymbol{W}_{\boldsymbol{x}_f}\left(\boldsymbol{x}(t_f) - \boldsymbol{x}_f\right) + \left(\dot{\boldsymbol{x}} - H_{\boldsymbol{\lambda}}\right) \\ \left(\dot{\boldsymbol{\lambda}} + H_{\boldsymbol{x}}\right) \\ \boldsymbol{p}_{\boldsymbol{u}} + \dot{\boldsymbol{u}}\dfrac{\delta t_f}{\delta \tau} \end{bmatrix}\right|_{t_f} \tag{32}$$

$$\frac{\delta t_f}{\delta \tau} = -k_{t_f} g(t_f) \tag{33}$$

where

$$\boldsymbol{p} = \begin{bmatrix} \boldsymbol{p}_{\boldsymbol{x}} \\ \boldsymbol{p}_{\boldsymbol{\lambda}} \\ \boldsymbol{p}_{\boldsymbol{u}} \end{bmatrix} = H_{\boldsymbol{yy}}\boldsymbol{v} + \boldsymbol{M}\dot{\boldsymbol{y}} - \begin{bmatrix} \ddot{\boldsymbol{x}} \\ \ddot{\boldsymbol{\lambda}} \\ \boldsymbol{0} \end{bmatrix} + (H+1)\begin{bmatrix} H_{\boldsymbol{x}} \\ \boldsymbol{f} \\ H_{\boldsymbol{u}} \end{bmatrix} \tag{34}$$

$$g = H_{\boldsymbol{x}}{}^{\mathrm{T}} H_{\boldsymbol{x}} + H_{\boldsymbol{\lambda}}{}^{\mathrm{T}} H_{\boldsymbol{\lambda}} - \dot{\boldsymbol{x}}^{\mathrm{T}}\dot{\boldsymbol{x}} - \dot{\boldsymbol{\lambda}}^{\mathrm{T}}\dot{\boldsymbol{\lambda}} + \left(H+1\right)^2 \tag{35}$$

$H_{\boldsymbol{yy}} = \begin{bmatrix} H_{\boldsymbol{xx}} & \boldsymbol{f}_{\boldsymbol{x}}{}^{\mathrm{T}} & H_{\boldsymbol{xu}} \\ \boldsymbol{f}_{\boldsymbol{x}} & \boldsymbol{0} & \boldsymbol{f}_{\boldsymbol{u}} \\ H_{\boldsymbol{ux}} & \boldsymbol{f}_{\boldsymbol{u}}{}^{\mathrm{T}} & H_{\boldsymbol{uu}} \end{bmatrix}$ is the Hessian matrix, $\boldsymbol{v} = \begin{bmatrix} \left(H_{\boldsymbol{x}} + \dot{\boldsymbol{\lambda}}\right) \\ \left(\boldsymbol{f} - \dot{\boldsymbol{x}}\right) \\ \boldsymbol{0} \end{bmatrix}$ is the optimality vector, the matrix $\boldsymbol{M}$ is $\boldsymbol{M} = \begin{bmatrix} \boldsymbol{f}_{\boldsymbol{x}} & \boldsymbol{0} & \boldsymbol{f}_{\boldsymbol{u}} \\ -H_{\boldsymbol{xx}} & -\boldsymbol{f}_{\boldsymbol{x}}{}^{\mathrm{T}} & -H_{\boldsymbol{xu}} \\ \boldsymbol{0} & \boldsymbol{0} & \boldsymbol{0} \end{bmatrix}$, $\ddot{\boldsymbol{x}} = \frac{\mathrm{d}^2\boldsymbol{x}}{\mathrm{d}t^2}$ and $\ddot{\boldsymbol{\lambda}} = \frac{\mathrm{d}^2\boldsymbol{\lambda}}{\mathrm{d}t^2}$. $\frac{\delta \boldsymbol{y}_f}{\delta \tau} = \frac{\delta \boldsymbol{y}(t_f)}{\delta \tau} + \dot{\boldsymbol{y}}\frac{\delta t_f}{\delta \tau}$ is the derivative of variation in the terminal variable with respect to $\tau$. $\boldsymbol{K}$ is a $(2n+m)\times(2n+m)$ dimensional positive definite matrix and $k_{t_f}$ is a positive constant.

**Theorem 1:** Solving the IVP with respect to $\tau$, defined by (30)-(33) from an arbitrarily initial solution, when $\tau \to +\infty$, we have $J_2 \to 0$ and $(\boldsymbol{x}, \boldsymbol{\lambda}, \boldsymbol{u})$ will satisfy the necessary optimality conditions of Problem 2.

Proof: Obviously, the minimum of the unconstrained functional defined in Problem 3 is $J_2 = 0$, which determines (17), (18), (20), (21) and (25). This means the optimal solution of functional (26) satisfies the optimality conditions of Problem 2. Differentiating (26) with respect to $\tau$ and substituting (30)-(33) in, we have $\frac{\delta J_2}{\delta \tau} \le 0$. The functional $J_2$ will decrease until $J_2 = 0$, which occurs when $\tau \to +\infty$. When $J_2$ reaches the minimum, i.e., $J_2 = 0$, this determines the necessary optimality conditions of Problem 2. ■

What needs to be pointed out is that the equilibrium point of the variation dynamic evolving equations (30)-(33) does not satisfy Eq. (22) necessarily. For example, the solution maybe meets

$$\boldsymbol{u} = \arg \min_{2u_{\min} \le u_i \le 2u_{\max}} (H) \tag{36}$$

Thus, when initializing the computation, the initial value $t_f\big|_{\tau=0}$ will be set small to drive the control variables to be large, and then we embody the control constraint with the limited integrator bounded by $[u_{\min}, u_{\max}]$ on $\boldsymbol{u}$ for the right solution. Moreover, Because of the intrinsic property of time-optimal control problems, when implementing the numerical computation, a crucial issue arises that $\boldsymbol{u}$ may be discontinuous, typically in bang-bang form, and $\boldsymbol{x}$, $\boldsymbol{\lambda}$ are non-continuously differentiable. Thus, equation (30) must be used cautiously during the discretization. Suppose that $\boldsymbol{x}$ and $\boldsymbol{\lambda}$ are piecewise continuously differentiable and $\boldsymbol{u}$ is piecewise continuous, we capture the control switches by introducing corner points to partition the functional. For example

$$\int_{t_0}^{t_f} F\mathrm{d}t = \int_{t_0}^{t_c^-} F\mathrm{d}t + \int_{t_c^+}^{t_f} F\mathrm{d}t \tag{37}$$

where $t_c^-$ and $t_c^+$ are the times right before and immediately after the corner time $t_c$. Through the variation theory, the corner optimality conditions are

$$\frac{\delta \boldsymbol{y}_{c^-}}{\delta \tau} = -2\boldsymbol{K}\begin{bmatrix} (\dot{\boldsymbol{x}} - H_{\boldsymbol{\lambda}})\big|_{t_c^-} - (\dot{\boldsymbol{x}} - H_{\boldsymbol{\lambda}})\big|_{t_c^+} \\ (\dot{\boldsymbol{\lambda}} + H_{\boldsymbol{x}})\big|_{t_c^-} - (\dot{\boldsymbol{\lambda}} + H_{\boldsymbol{x}})\big|_{t_c^+} \\ \boldsymbol{p}_{\boldsymbol{u}}(t_c^-) + \dot{\boldsymbol{u}}(t_c^-)\dfrac{\delta t_c}{\delta \tau} \end{bmatrix} \tag{38}$$

$$\frac{\delta \boldsymbol{y}_{c^+}}{\delta \tau} = -2\boldsymbol{K}\begin{bmatrix} (\dot{\boldsymbol{x}} - H_{\boldsymbol{\lambda}})\big|_{t_c^-} - (\dot{\boldsymbol{x}} - H_{\boldsymbol{\lambda}})\big|_{t_c^+} \\ (\dot{\boldsymbol{\lambda}} + H_{\boldsymbol{x}})\big|_{t_c^-} - (\dot{\boldsymbol{\lambda}} + H_{\boldsymbol{x}})\big|_{t_c^+} \\ \boldsymbol{p}_{\boldsymbol{u}}(t_c^+) + \dot{\boldsymbol{u}}(t_c^+)\dfrac{\delta t_c}{\delta \tau} \end{bmatrix} \tag{39}$$

$$\frac{\delta t_c}{\delta \tau} = -k_{t_c}\left(g(t_c^-) - g(t_c^+)\right) \tag{40}$$

where $\dfrac{\delta \boldsymbol{y}_{c^-}}{\delta \tau} = \dfrac{\delta \boldsymbol{y}(t_c^-)}{\delta \tau} + \dot{\boldsymbol{y}}(t_c^-)\dfrac{\delta t_c}{\delta \tau}$, $\dfrac{\delta \boldsymbol{y}_{c^+}}{\delta \tau} = \dfrac{\delta \boldsymbol{y}(t_c^+)}{\delta \tau} + \dot{\boldsymbol{y}}(t_c^+)\dfrac{\delta t_c}{\delta \tau}$, and $k_{t_c}$ is a positive constant.

**Remark 1**: If in Problem 2 the terminal boundary conditions for the states $\boldsymbol{x} = \begin{bmatrix} \boldsymbol{x}_1 \\ \boldsymbol{x}_2 \end{bmatrix}$ are only partly fixed, as $\boldsymbol{x}_1(t_f) = \boldsymbol{x}_{1f}$ while $\boldsymbol{x}_2(t_f)$ is free, then the transversality conditions that determines the optimal solution include

$$\boldsymbol{\lambda}_2(t_f) = \boldsymbol{0} \tag{41}$$

where $\boldsymbol{\lambda} = \begin{bmatrix} \boldsymbol{\lambda}_1 \\ \boldsymbol{\lambda}_2 \end{bmatrix}$ is the corresponding costate. Using the VEM, the equivalent unconstrained functional may be constructed as

$$\begin{aligned} J_2 = & (\boldsymbol{x}(t_0) - \boldsymbol{x}_0)^{\mathrm{T}} \boldsymbol{W}_{\boldsymbol{x}_0} (\boldsymbol{x}(t_0) - \boldsymbol{x}_0) + \left(\boldsymbol{x}_1(t_f) - \boldsymbol{x}_{1f}\right)^{\mathrm{T}} \boldsymbol{W}_{\boldsymbol{x}_{1f}} \left(\boldsymbol{x}_1(t_f) - \boldsymbol{x}_{1f}\right) + \boldsymbol{\lambda}_2(t_f)^{\mathrm{T}} \boldsymbol{W}_{\boldsymbol{\lambda}_{2f}} \boldsymbol{\lambda}_2(t_f) \\ & + \int_{t_0}^{t_f} \left\{ (\dot{\boldsymbol{x}} - H_{\boldsymbol{\lambda}})^{\mathrm{T}} (\dot{\boldsymbol{x}} - H_{\boldsymbol{\lambda}}) + (\dot{\boldsymbol{\lambda}} + H_{\boldsymbol{x}})^{\mathrm{T}} (\dot{\boldsymbol{\lambda}} + H_{\boldsymbol{x}}) + (H+1)^2 \right\} \mathrm{d}t \end{aligned} \tag{42}$$

where $\boldsymbol{W}_{\boldsymbol{x}_0}$, $\boldsymbol{W}_{\boldsymbol{x}_{1f}}$ and $\boldsymbol{W}_{\boldsymbol{\lambda}_{2f}}$ are right-dimensional weighted positive definite matrixes, and the variation dynamic evolving equations deduced is similar to the all terminal states fixed case except (32) is reformulated as

$$\frac{\delta \boldsymbol{y}_f}{\delta \tau} = -2\boldsymbol{K}\begin{bmatrix} \boldsymbol{W}_{\boldsymbol{x}_{1f}}\left(\boldsymbol{x}_1(t_f)-\boldsymbol{x}_{1f}\right)+\left(\dot{\boldsymbol{x}}_1 - H_{\boldsymbol{\lambda}_1}\right) \\ \left(\dot{\boldsymbol{x}}_2 - H_{\boldsymbol{\lambda}_2}\right) \\ \left(\dot{\boldsymbol{\lambda}}_1 + H_{\boldsymbol{x}_1}\right) \\ \boldsymbol{W}_{\boldsymbol{\lambda}_{2f}}\boldsymbol{\lambda}_2(t_f)+\left(\dot{\boldsymbol{\lambda}}_2 + H_{\boldsymbol{x}_2}\right) \\ \boldsymbol{p}_{\boldsymbol{u}} + \dot{\boldsymbol{u}}\dfrac{\delta t_f}{\delta \tau} \end{bmatrix}_{t_f} \tag{43}$$

**Remark 2**: If in Problem 2 the terminal boundary conditions for the states are generally constrained by a $s$-dimensional ($s \le n$) vector function as $\boldsymbol{\xi}\left(\boldsymbol{x}(t_f)\right)=\boldsymbol{0}$, then the transversality conditions that determines the optimal solution include

$$\boldsymbol{\lambda}(t_f)=\left(\xi_{\boldsymbol{x}(t_f)}\right)^{\mathrm{T}}\boldsymbol{\pi} \tag{44}$$

where $\boldsymbol{\pi}$ is the Lagrange multiplier adjoined with the terminal boundary conditions in deriving the optimality condition, under the frame of the adjoining method [15]. Using the VEM, the equivalent unconstrained functional may be constructed as

$$\begin{aligned} J_2 =& \left(\boldsymbol{x}(t_0)-\boldsymbol{x}_0\right)^{\mathrm{T}}\boldsymbol{W}_{\boldsymbol{x}_0}\left(\boldsymbol{x}(t_0)-\boldsymbol{x}_0\right)+\xi\left(\boldsymbol{x}(t_f)\right)^{\mathrm{T}}\boldsymbol{W}_{\boldsymbol{x}_f}\xi\left(\boldsymbol{x}(t_f)\right) \\ &+\left(\boldsymbol{\lambda}(t_f)-(\xi_{\boldsymbol{x}(t_f)})^{\mathrm{T}}\boldsymbol{\pi}\right)^{\mathrm{T}}\boldsymbol{W}_{\boldsymbol{\lambda}_f}\left(\boldsymbol{\lambda}(t_f)-(\xi_{\boldsymbol{x}(t_f)})^{\mathrm{T}}\boldsymbol{\pi}\right) \\ &+\int_{t_0}^{t_f}\left\{\left(\dot{\boldsymbol{x}}-H_{\boldsymbol{\lambda}}\right)^{\mathrm{T}}\left(\dot{\boldsymbol{x}}-H_{\boldsymbol{\lambda}}\right)+\left(\dot{\boldsymbol{\lambda}}+H_{\boldsymbol{x}}\right)^{\mathrm{T}}\left(\dot{\boldsymbol{\lambda}}+H_{\boldsymbol{x}}\right)+\left(H+1\right)^2\right\}\mathrm{d}t \end{aligned} \tag{45}$$

where $\boldsymbol{W}_{\boldsymbol{x}_0}$, $\boldsymbol{W}_{\boldsymbol{x}_f}$ and $\boldsymbol{W}_{\boldsymbol{\lambda}_f}$ are right-dimensional weighted positive definite matrixes, and the variation dynamic evolving equations deduced is similar to the all terminal states fixed case except Eq. (32) is reformulated as

$$\frac{\delta \boldsymbol{y}_f}{\delta \tau} = -2\boldsymbol{K}\begin{bmatrix} \xi_{\boldsymbol{x}(t_f)}{}^{\mathrm{T}}\boldsymbol{W}_{\boldsymbol{x}_f}\xi+\left(\dot{\boldsymbol{x}}-H_{\boldsymbol{\lambda}}\right) \\ \boldsymbol{W}_{\boldsymbol{\lambda}_f}\left(\boldsymbol{\lambda}(t_f)-(\xi_{\boldsymbol{x}(t_f)})^{\mathrm{T}}\boldsymbol{\pi}\right)+\left(\dot{\boldsymbol{\lambda}}+H_{\boldsymbol{x}}\right) \\ \boldsymbol{p}_{\boldsymbol{u}}+\dot{\boldsymbol{u}}\dfrac{\delta t_f}{\delta \tau} \end{bmatrix}_{t_f} \tag{46}$$

and new equations regarding the parameters $\boldsymbol{\pi}$ are introduced as

$$\frac{\delta \boldsymbol{\pi}}{\delta \tau} = \boldsymbol{K}_{\boldsymbol{\pi}}\xi_{\boldsymbol{x}(t_f)}\boldsymbol{W}_{\boldsymbol{\lambda}_f}\left(\boldsymbol{\lambda}(t_f)-(\xi_{\boldsymbol{x}(t_f)})^{\mathrm{T}}\boldsymbol{\pi}\right) \tag{47}$$

where $\boldsymbol{K}_{\boldsymbol{\pi}}$ is a $s \times s$ dimensional positive definite matrix.

Similarly, we may use the partial differential operator "$\partial$" and the differential operator "$\mathrm{d}$" to reformulate the variation dynamic evolving equations (30)-(33) to get the following Evolution PDE (EPDE) and Evolution Differential Equations (EDEs) as

$$\frac{\partial}{\partial \tau}\begin{bmatrix}\boldsymbol{x}\\ \boldsymbol{\lambda}\\ \boldsymbol{u}\end{bmatrix} = -2\boldsymbol{K}\left( H_{\boldsymbol{yy}}\begin{bmatrix}\left(H_{\boldsymbol{x}}+\dfrac{\partial \boldsymbol{\lambda}}{\partial t}\right)\\ \left(\boldsymbol{f}-\dfrac{\partial \boldsymbol{x}}{\partial t}\right)\\ \boldsymbol{0}\end{bmatrix} - \frac{\partial}{\partial t}\begin{bmatrix}\left(\dfrac{\partial \boldsymbol{x}}{\partial t}-\boldsymbol{f}\right)\\ \left(\dfrac{\partial \boldsymbol{\lambda}}{\partial t}+H_{\boldsymbol{x}}\right)\\ \boldsymbol{0}\end{bmatrix} + (H+1)\begin{bmatrix}H_{\boldsymbol{x}}\\ \boldsymbol{f}\\ H_{\boldsymbol{u}}\end{bmatrix}\right), \quad t\in(t_0,t_f) \tag{48}$$

$$\frac{\mathrm{d}\boldsymbol{y}(t_0)}{\mathrm{d}\tau} = 2\boldsymbol{K}\begin{bmatrix} \boldsymbol{W}_{\boldsymbol{x}_0}\left(\boldsymbol{x}_0 - \boldsymbol{x}(t_0)\right) + \left(\frac{\partial \boldsymbol{x}}{\partial t} - H_{\boldsymbol{\lambda}}\right) \\ \left(\frac{\partial \boldsymbol{\lambda}}{\partial t} + H_{\boldsymbol{x}}\right) \\ \boldsymbol{p}_{\boldsymbol{u}} \end{bmatrix}\Bigg|_{t_0} \tag{49}$$

$$\frac{\mathrm{d}\boldsymbol{y}_f}{\mathrm{d}\tau} = -2\boldsymbol{K}\begin{bmatrix} \boldsymbol{W}_{\boldsymbol{x}_f}\left(\boldsymbol{x}(t_f) - \boldsymbol{x}_f\right) + \left(\frac{\partial \boldsymbol{x}}{\partial t} - H_{\boldsymbol{\lambda}}\right) \\ \left(\frac{\partial \boldsymbol{\lambda}}{\partial t} + H_{\boldsymbol{x}}\right) \\ \boldsymbol{p}_{\boldsymbol{u}} + \frac{\partial \boldsymbol{u}}{\partial t}\frac{\mathrm{d}t_f}{\mathrm{d}\tau} \end{bmatrix}\Bigg|_{t_f} \tag{50}$$

$$\frac{\mathrm{d}t_f}{\mathrm{d}\tau} = -k_{t_f}\, g(t_f) \tag{51}$$

Put into this perspective, the definite conditions are the initial guess of $t_f$, i.e., $t_f\big|_{\tau=0}$, and

$$\begin{bmatrix} \boldsymbol{x}(t,\tau) \\ \boldsymbol{\lambda}(t,\tau) \\ \boldsymbol{u}(t,\tau) \end{bmatrix}\Bigg|_{\tau=0} = \begin{bmatrix} \tilde{\boldsymbol{x}}(t) \\ \tilde{\boldsymbol{\lambda}}(t) \\ \tilde{\boldsymbol{u}}(t) \end{bmatrix} \tag{52}$$

where $\tilde{\boldsymbol{x}}(t)$, $\tilde{\boldsymbol{\lambda}}(t)$ and $\tilde{\boldsymbol{u}}(t)$ are the initial guesses of the variables. Recall Fig. 2, Eqs. (48)-(51) realize the anticipated variable evolving along the variation time $\tau$. The initial conditions of $\boldsymbol{x}(t,\tau)$, $\boldsymbol{\lambda}(t,\tau)$ and $\boldsymbol{u}(t,\tau)$ at $\tau = 0$ are arbitrary and their value at $\tau = +\infty$ may be the optimal solution of the OCP. The right part of the EPDE (48) is also only a vector function of time $t$. Thus we may apply the semi-discrete method to discretize it along the normal time dimension and further use ODE integration methods to get the numerical solution.

## V. An Illustrative Example

The strategy is tested by applying to a linear example with analytic solution [19]. However, its application is not restricted to the linear case. Consider the following dynamic system

$$\dot{\boldsymbol{x}} = \boldsymbol{A}\boldsymbol{x} + \boldsymbol{b}u$$

where $\boldsymbol{x} = \begin{bmatrix} x_1 \\ x_2 \end{bmatrix}$, $\boldsymbol{A} = \begin{bmatrix} 0 & 1 \\ 0 & 0 \end{bmatrix}$, $\boldsymbol{b} = \begin{bmatrix} 0 \\ 1 \end{bmatrix}$. Find the solution that minimizes the performance index (14) with the control constraint $-1 \le u \le 1$ and boundary conditions

$$\boldsymbol{x}(t_0) = \begin{bmatrix} 1 \\ 1 \end{bmatrix}, \boldsymbol{x}(t_f) = \begin{bmatrix} 0 \\ 0 \end{bmatrix}$$

where the initial time $t_0 = 0$ is fixed.

In solving this example using the VEM, the EPDE derived is

$$\frac{\partial}{\partial \tau}\begin{bmatrix} \boldsymbol{x} \\ \boldsymbol{\lambda} \\ \boldsymbol{u} \end{bmatrix} = -2\boldsymbol{K}\left( \begin{bmatrix} \boldsymbol{0} & \boldsymbol{A}^{\mathrm{T}} & \boldsymbol{0} \\ \boldsymbol{A} & \boldsymbol{0} & \boldsymbol{b} \\ \boldsymbol{0} & \boldsymbol{b}^{\mathrm{T}} & 1 \end{bmatrix} \begin{bmatrix} \left( \boldsymbol{A}^{\mathrm{T}}\boldsymbol{\lambda} + \frac{\partial \boldsymbol{\lambda}}{\partial t} \right) \\ \left( \boldsymbol{A}\boldsymbol{x} + \boldsymbol{b}u - \frac{\partial \boldsymbol{x}}{\partial t} \right) \\ \boldsymbol{0} \end{bmatrix} - \frac{\partial}{\partial t}\begin{bmatrix} \left( \frac{\partial \boldsymbol{x}}{\partial t} - \boldsymbol{A}\boldsymbol{x} - \boldsymbol{b}u \right) \\ \left( \frac{\partial \boldsymbol{\lambda}}{\partial t} + \boldsymbol{A}^{\mathrm{T}}\boldsymbol{\lambda} \right) \\ \boldsymbol{0} \end{bmatrix} + (\boldsymbol{\lambda}^{\mathrm{T}}\boldsymbol{A}\boldsymbol{x} + \boldsymbol{\lambda}^{\mathrm{T}}\boldsymbol{b}u + 1)\begin{bmatrix} \boldsymbol{A}^{\mathrm{T}}\boldsymbol{\lambda} \\ \boldsymbol{A}\boldsymbol{x} + \boldsymbol{b}u \\ u + \boldsymbol{\lambda}^{\mathrm{T}}\boldsymbol{b} \end{bmatrix} \right)$$

where $\boldsymbol{K} = 30 \cdot \boldsymbol{I}_{5\times5}$ with $\boldsymbol{I}_{5\times5}$ being $5\times5$ dimensional identity matrix. The specific EDEs may be obtained accordingly with weighted matrixes $\boldsymbol{W}_{\boldsymbol{x}_0} = \boldsymbol{I}_{2\times2}$ and $\boldsymbol{W}_{\boldsymbol{x}_f} = \boldsymbol{I}_{2\times2}$. The parameter $k_{t_f}$ is $k_{t_f} = 30$. An initial value of the terminal time $t_f\big|_{\tau=0} = 2$ was set and the definite conditions of the EPDE, i.e., the initial guess of the variables $\tilde{\boldsymbol{x}}(t)$, $\tilde{\boldsymbol{\lambda}}(t)$ and $\tilde{u}(t)$, were given by $\tilde{\boldsymbol{x}}(t) = \begin{bmatrix} 0.3 \\ 0.5 \end{bmatrix}$, $\tilde{\boldsymbol{\lambda}}(t) = \begin{bmatrix} 0 \\ 0 \end{bmatrix}$, and $\tilde{u}(t) = 0$. We discretized the time horizon $[t_0, t_f]$ uniformly, with 41 points. Especially, one corner point, with initial guess $t_c\big|_{\tau=0} = 1$ s, was also added to the discretization points to capture the occurrence of discontinuity. Thus, a large IVP with 217 states (including the terminal time, the corner time and the variables at corner time point) was obtained. Since we did not specially scale the problem, we employed a stiff ODE integration method, "ode15s" in Matlab, for the numerical integration. In the integrator setting, the default relative error tolerance and the absolute error tolerance are $1\times10^{-3}$ and $1\times10^{-6}$, respectively.

For comparison, the analytic solution of this example is presented.

$$t \in [0,\, 1+(\sqrt{6}/2)) \qquad \begin{cases} \hat{x}_1 = -0.5t^2 + t + 1 \\ \hat{x}_2 = -t + 1 \\ \hat{\lambda}_1 = \sqrt{6}/3 \\ \hat{\lambda}_2 = -(\sqrt{6}/3)t + (\sqrt{6}/3) + 1 \\ \hat{u} = -1 \end{cases}$$

$$t \in [1+(\sqrt{6}/2),\, 1+\sqrt{6}] \qquad \begin{cases} \hat{x}_1 = 0.5t^2 - (1+\sqrt{6})t + 3.5 + \sqrt{6} \\ \hat{x}_2 = t - 1 - \sqrt{6} \\ \hat{\lambda}_1 = \sqrt{6}/3 \\ \hat{\lambda}_2 = -(\sqrt{6}/3)t + (\sqrt{6}/3) + 1 \\ \hat{u} = 1 \end{cases}$$

Figs. 3 and 4 show the evolving process of $x_1$ and $\lambda_1$ solutions to the optimal, respectively. The control results are plotted in Fig. 5. They show the asymptotically approach of the numerical results to the optimal. At $\tau$ = 600s, the numerical solutions are indistinguishable from the analytic, and it is found that the control switch is accurately captured. Fig. 6 plots the numerical solutions of $x_2$ and $\lambda_2$ at $\tau$ = 600s. They are almost identical with the analytic solutions. Especially, the sharp angle in the curve of $x_2$ is clearly shown. Described in the state plane, the states results are again compared with the analytic solution in Fig. 7. It illustrates the evolution process of the states, from an arbitrarily set initial point to the optimal results. The profiles for the terminal time and the corner time are presented in Fig 8. At $\tau$ = 600s, we compute that $t_f$ = 3.4492s and $t_c$ = 2.2249. They are very close to the analytic results.

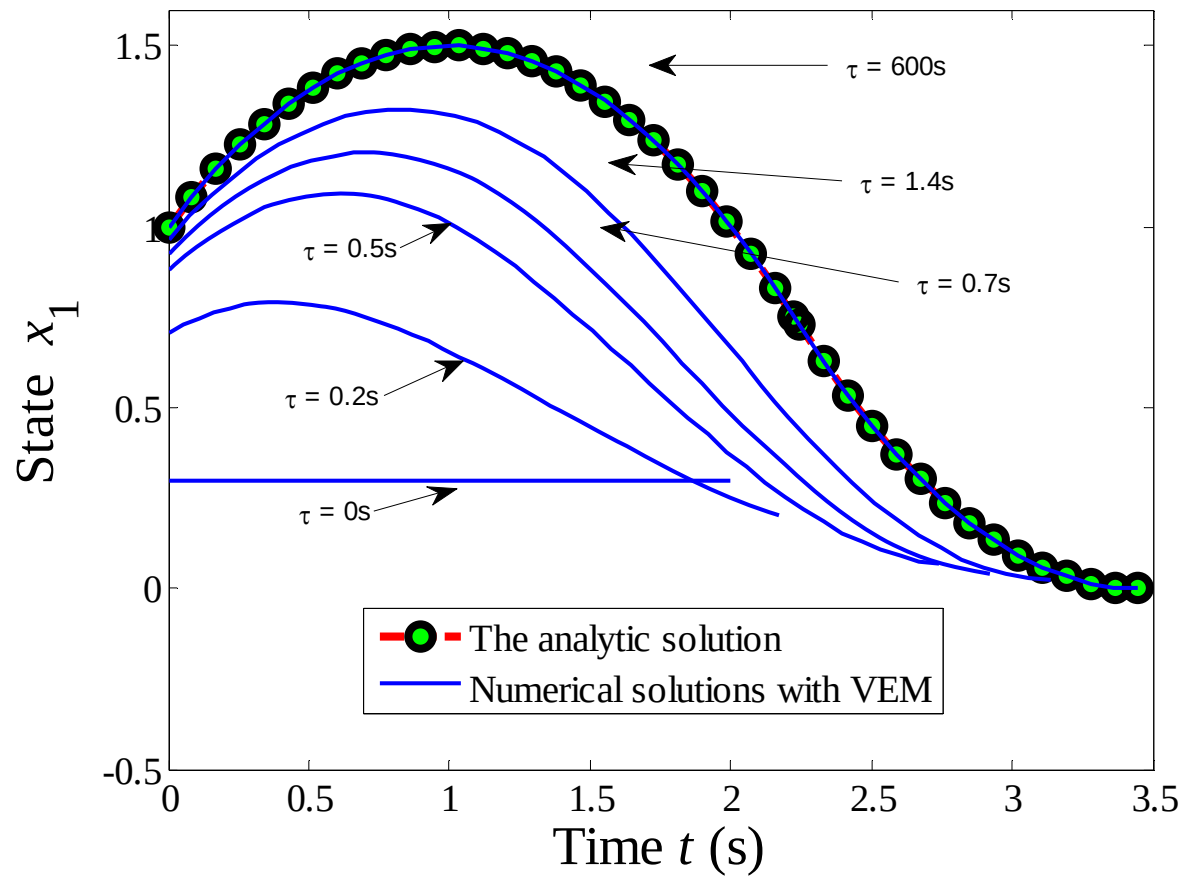

Fig. 3. The evolving of numerical solutions of $x_1$ to the analytic solution.

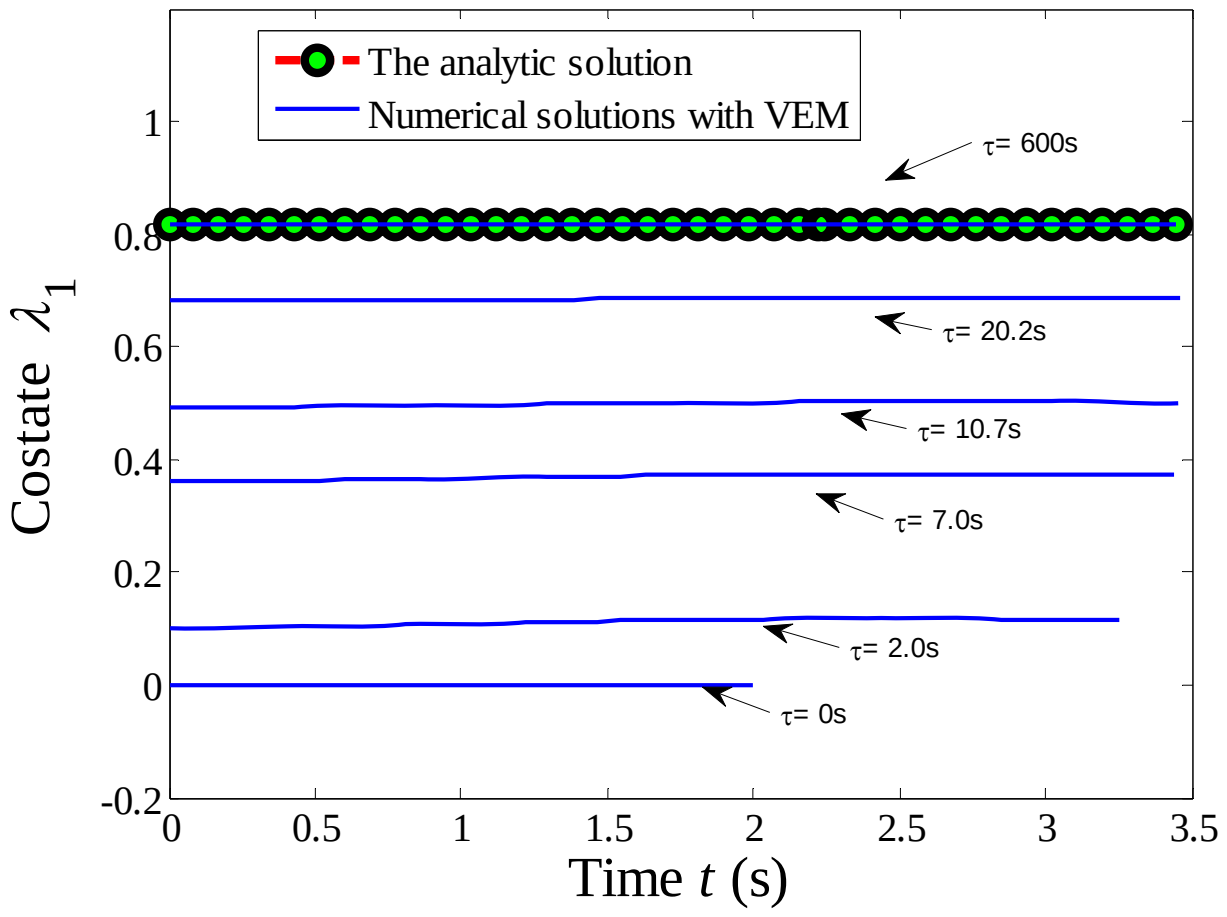

Fig. 4. The evolving of numerical solutions of $\lambda_1$ to the analytic solution.

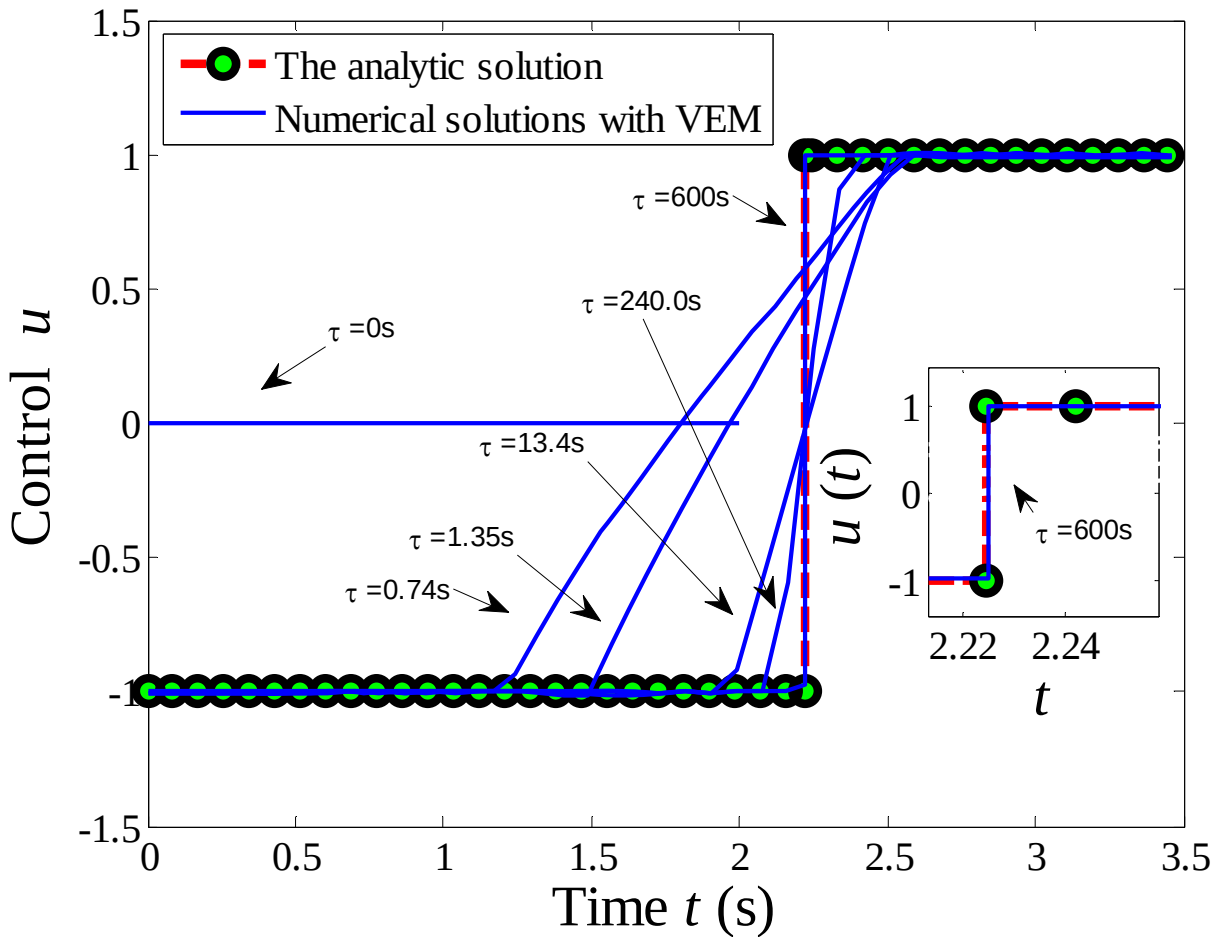

Fig. 5. The evolving of numerical solutions of $u$ to the analytic solution.

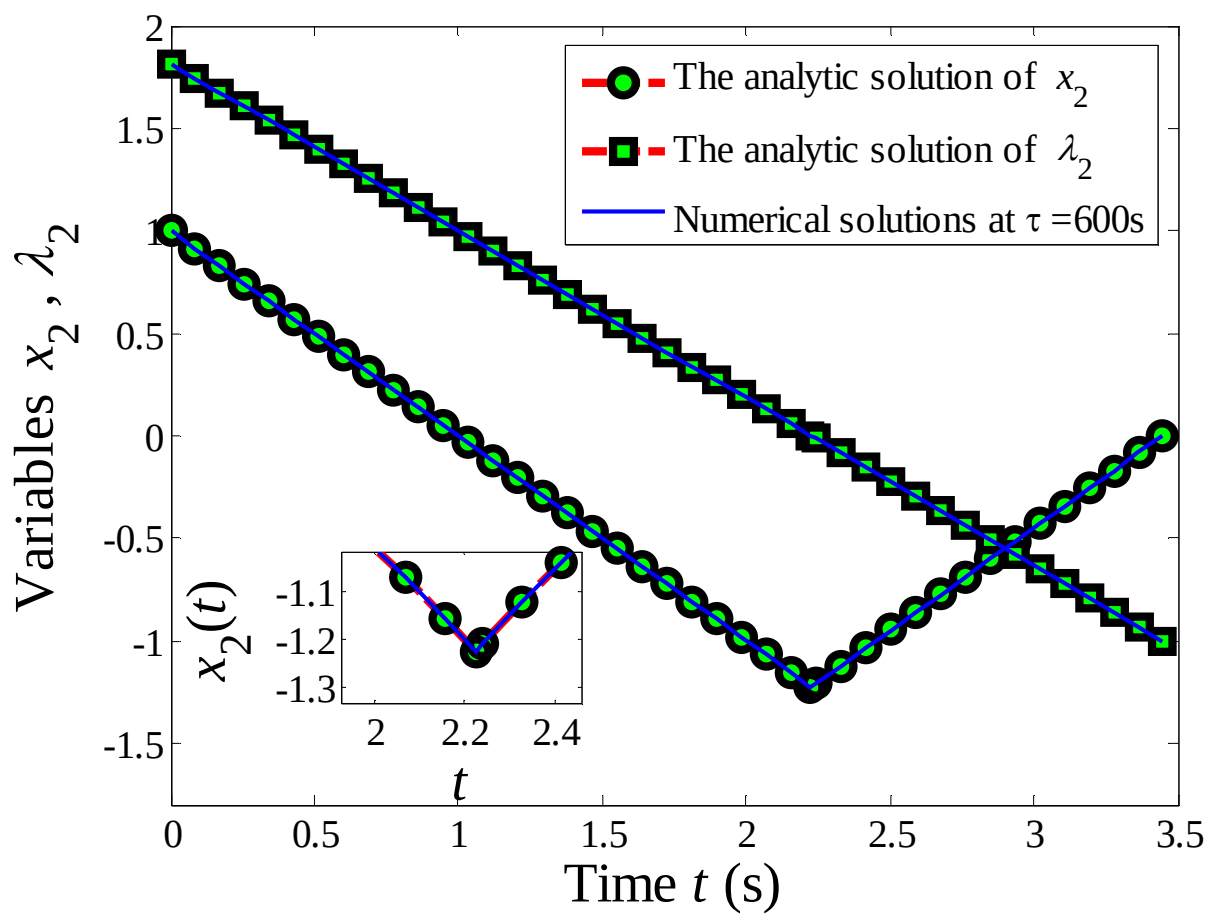

Fig. 6. The numerical solutions of $x_2$ and $\lambda_2$ with the VEM at $\tau = 600\text{s}$ .

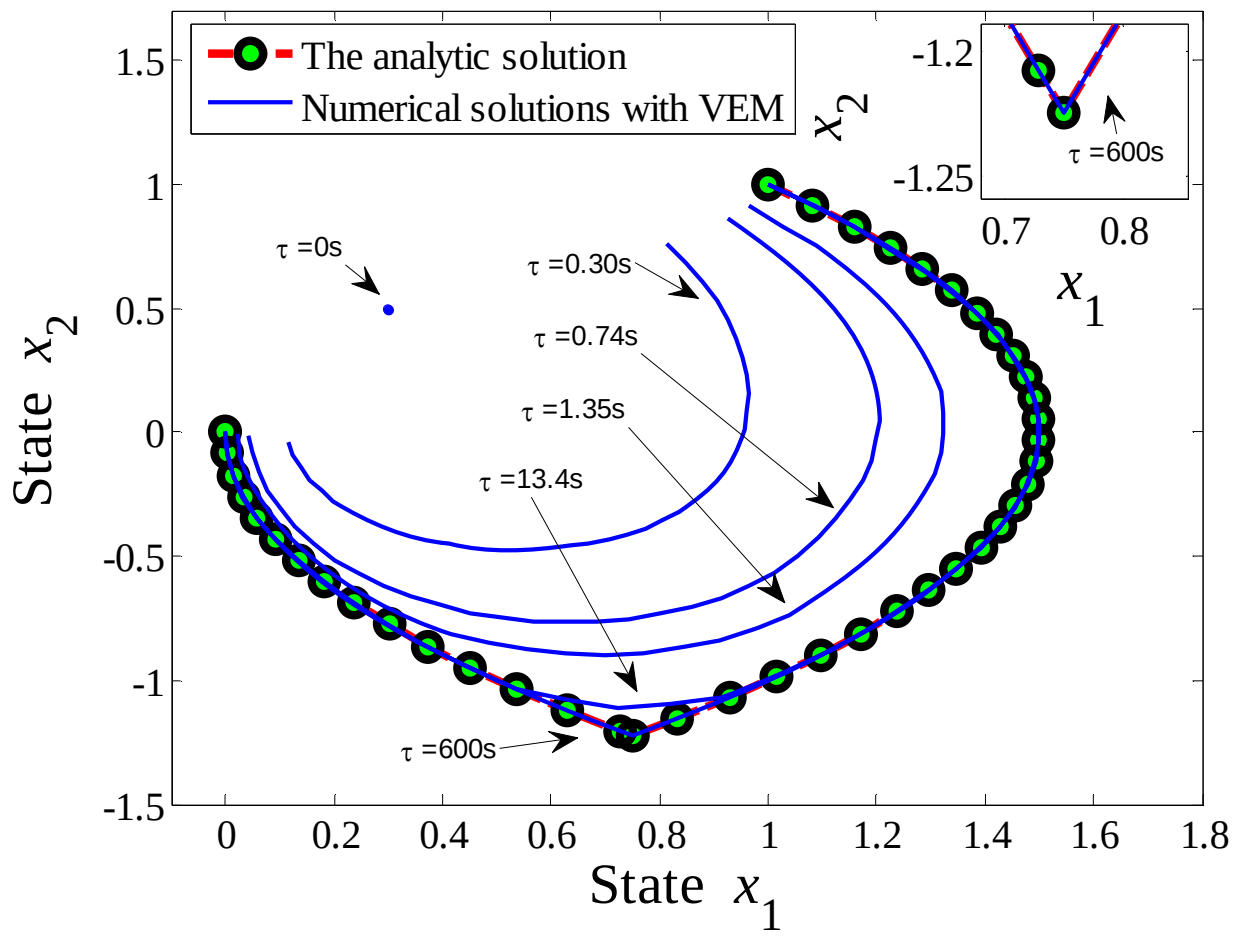

Fig. 7. The evolving of numerical solutions in $x_1 x_2$ state plane to the analytic solution.

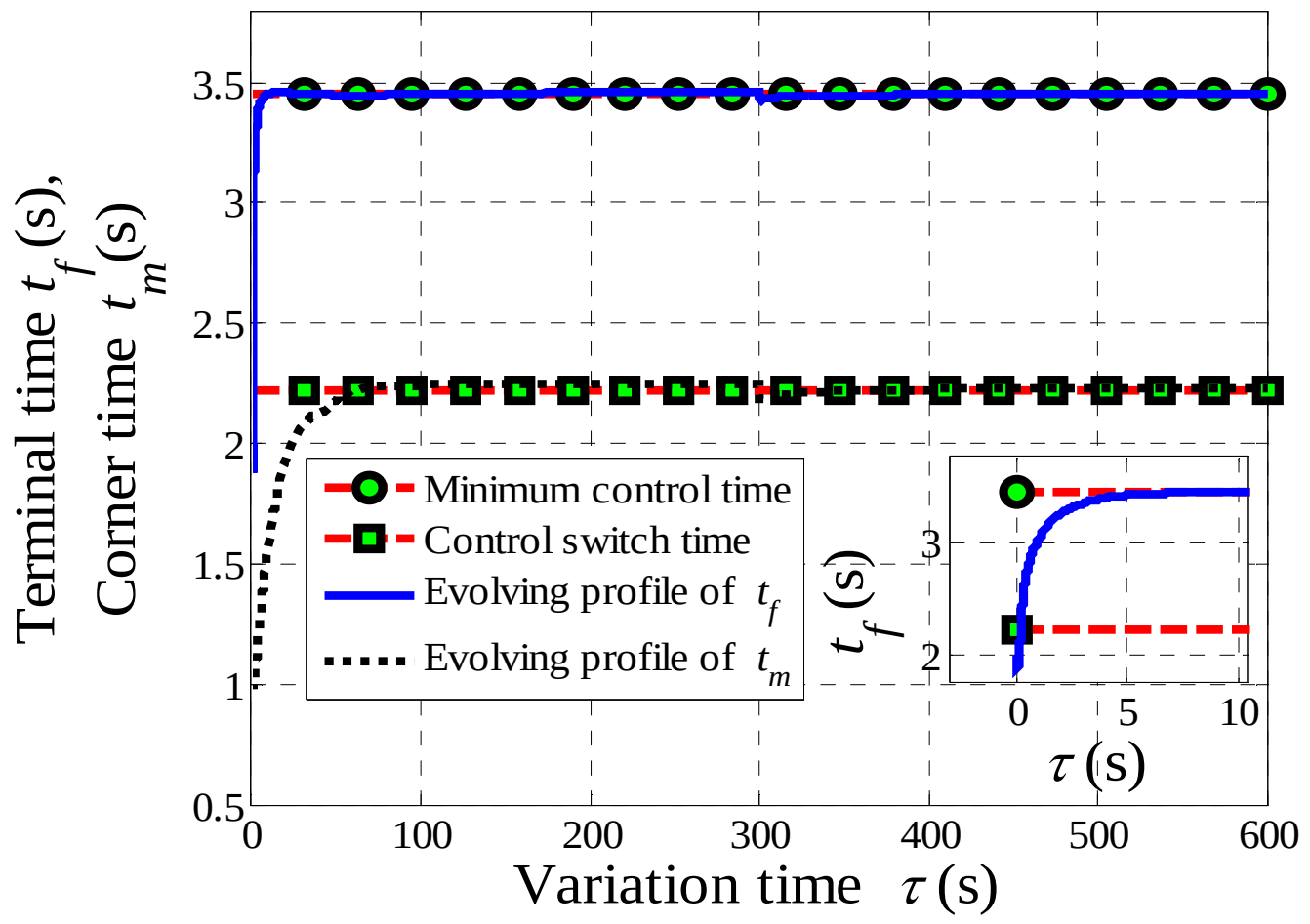

Fig. 8. The evolving profiles of $t_f$ and $t_c$ to the analytic results.

## VI. CONCLUSION

In this paper, some explosive work towards the computation of the constrained Optimal Control Problem (OCP), based on the variation evolution principle, is carried out, and an effective form of the Variation Evolving Method (VEM) is developed for the classic time-optimal control problem with control constraint. It is shown that the bang-bang structure and switch point of the optimal control could be accurately captured with the proposed method. In particular, the VEM is further developed in treating the OCP boundary conditions, and arbitrary initial values of variables could be used for the resulting Initial-value Problem (IVP). This treatment is also applicable to the work in Ref. [11] and it brings extra flexibility for the computation.